# Decentralized Spectrum Learning for IoT Wireless Networks Collision Mitigation

Christophe Moy
Univ Rennes
CNRS, IETR - UMR 6164
F-35000, Rennes, France,
christophe.moy@univ-rennes1.fr

Lilian Besson
CentraleSupélec
CNRS, IETR - UMR 6164
F-35576, Cesson-Sévigné, France
lilian.besson@centralesupelec.fr

*Abstract*—This paper describes the principles and implementation results of reinforcement learning algorithms on IoT devices for radio collision mitigation in ISM unlicensed bands. Learning is here used to improve both the IoT network capability to support a larger number of objects as well as the autonomy of IoT devices. We first illustrate the efficiency of the proposed approach in a proof-of-concept based on USRP software radio platforms operating on real radio signals. It shows how collisions with other RF signals present in the ISM band are diminished for a given IoT device. Then we describe the first implementation of learning algorithms on LoRa devices operating in a real LoRaWAN network, that we named IoTligent. The proposed solution adds neither processing overhead so that it can be ran in the IoT devices, nor network overhead so that no change is required to LoRaWAN. Real life experiments have been done in a realistic LoRa network and they show that IoTligent device battery life can be extended by a factor 2 in the scenarios we faced during our experiment.

## I. INTRODUCTION

Future Internet of Things (IoT) networks are expected to be used all around the world by thousands of devices with various wireless standards and abilities. In wireless IoT networks, and in particular in Low Power Wide Area Networks (LPWAN) operating in the unlicensed bands, the spectrum is shared by many end-devices without any coordination, neither between devices, nor between LPWAN networks themselves. This provokes collisions that could dramatically limit the promises expected for IoT applications. As a consequence, IoT wireless networks require to move towards smarter decentralized frequency resource allocation solutions. However, what will happen is hardly predictable at the expected scale of IoT devices life, which should be up to ten years. Due to both their extremely low cost and complexity requirements, IoT devices require distributed computational and energy efficient solutions that operate without any prior information, and that can deal with uncertainty. The aim of this article is to assess the potential benefit of reinforcement learning (RL) and especially of the Multi-Armed Bandit (MAB) framework, as a solution to the frequency allocation challenges arising in IoT networks.

We propose to show in this paper the theoretical foundation of this approach, and prove its viability at two levels in order to make IoT devices mitigate collisions with other Radio Frequency (RF) signals present in the ISM (Industrial Scientific and Medical) unlicensed band. The first step consists in a PoC (Proof-of-Concept) with a demonstrator running on real radio signals in laboratory conditions. Software Defined Radio (SDR) platforms are used here, and as far as we know it is the first implementation on real radio signals of learning running on the IoT devices side. The second step constitutes the first implementation of learning algorithms on devices deployed in a real LoRa network. The proposed implementation runs in the 868 MHz band, but could be used in any other ISM band, whatever the country. Any other IoT LPWAN (Low Power Wide Area Network) standard than LoRa could be targeted, as soon as channel assignation is not imposed by a central node in the network. We named *IoTligent* this decentralized (i.e., on device) and uncoordinated (devices do not communicate with each other) learning approach.

The rest of the article is organized as follows. The next section exposes the issue we target in this work and the corresponding hypotheses. Section II reminds the foundation of learning algorithms used in our appraoch. Then Section III details the IoT proof-of-concept design that has been done with USRP platforms, named *MALIN*. Section IV explains how this has been done in a real LoRa network using Pycom devices. Implementation details are given in Section V and finally Section VI gives results of experiments done in a real deployed LoRa network in the city of Rennes, France.

## I. ISSUE, HYPOTHESES AND DECENTRALIZATION PROS

### A. Collisions vs autonomy

The possibility of suffering from collisions is the main drawback of IoT in terms of battery autonomy at the first level, but also of IoT viability itself in the ISM bands. Indeed collisions may cause (many) retransmissions at the cost of an increase of the RF contention, and may lead to a lower battery lifetime. Even worse, this could derive to a total failure of the IoT device, either because it cannot succeed in sending data to the network, or because multiple repetitions could make it consume all its energy.

### B. Analysis of collisions

Radio collisions will be the weak point of LPWAN IoT networks operating in the unlicensed bands, such as ISM bands. Collisions may occur with:

- other IoT devices of the same network, as several networks covering the same area are not coordinated. This can occur between IoT devices uplink (UL) transmissions, and between IoT UL and gateway downlink (DL) transmissions towards IoT devices.

- Other IoT devices of surrounding networks others than the network of our device, but using the same IoT standard. This can occur both in UL and DL, as surrounding IoT gateways of different networks are *not* coordinated. They could use the same channels, or partly same and partly different channels.
- Other IoT radio signals using other IoT radio standards. Each IoT standard uses its own rules for channeling, bandwidth, user repartition, etc.
- Other radio signals present in the ISM band which are not IoT signals. By definition, they use completely different rules than IoT. They are "jammers" from the IoT networks point of view.

It is also important to note that each IoT standard indeed uses its own rules for channeling, bandwidth, user repartition, etc. Therefore, all this leads to an heratic use of spectrum that cannot be planned, and has to be learnt « in vivo ». However, unlicensed band does not mean un-ruled band (duty cycle, power limit, etc), but they are more exposed to the non-respect of these few rules as regulation is relaxed and thus, controls as well.

*C. A device-side solution for spectrum management*

Our learning approach imposes no change on a normal IoT protocol (as for instance LoRaWAN [1]): no extra retransmission, no extra-power to be sent, no data to be added in frames, etc. The only condition is that the proposed solution should work with the acknowledged mode for IoT. The underlying hypothesis is that "channels" (there are no official channels in ISM bands) occupancy by surrounding radio signals (IoT or not) is not equally balanced. In other words, some ISM sub-bands are less occupied or jammed than others, but it is not possible to predict it in time and space, so it has to be learnt on the fly.

The considered learning algorithm is a kind of artificial intelligence (AI) algorithm that is compatible with the IoT device low complexity, as we explain below. It is indeed much more efficient to implement radio collision mitigation approaches on the device side, as devices may be quite far away from gateways, and suffer from different radio and jamming/co-existence conditions. But they are the place where every Watt counts at transmission, and where sensitivity should be the best at reception, as no extra-processing can be afforded.

*D. Advantages of the proposed solution*

The proposed approach is based on reinforcement learning algorithms such as those already studied [2] and experimented on real radio signals for Cognitive Radio and especially Opportunistic Spectrum Access (OSA) [5]. We assert that, as for OSA, the IoT spectrum access issue can be modeled as a Multi-Armed Bandit (MAB) problem. Reinforcement learning is based on a feedback loop that gives a success measure of experience. In the IoT context, we propose to use the acknowledgement (ACK) sent by the gateway to the IoT device as a binary reward. Every device aims at maximizing its transmission success rate, or equivalently, at maximizing its cumulated reward.

The main advantages of our solution are:

- this algorithm has mathematical proofs of convergence,
- proofs are verified in real radio conditions, thanks to the good matching between the model and reality,
- learning converges effectively very fast in real experiments, thus it is adequate for radio applications [3],
- implementation and execution both require very low processing and memory overhead, so that it is possible to add the proposed approach in IoT devices for a negligible money cost, negligible complexity (processing, hardware, memory) and no extra-energy consumption overhead,
- learning starts from scratch, so there is no need for any prior knowledge at the beginning (and loose some time to acquire this knowledge),
- using such learning algorithm will never give worse results than a state-of-the-art random solution, even before learning brings some advantage, for instance at the very beginning of the learning process [6].

Hence we argue that the proposed approach can adapt to any kind of radio context, and we also note that:

- the stationarity of the environment is a requirement for the proofs of convergence, but if conditions change, convergence is so fast that a simple solution consist in reseting learning from time to time [6] (note that there also exist adaptive versions),
- no coordination is required between devices, but benefits decrease with the number of devices using the proposed solution, when it represents a great majority of devices (see solutions in [6][7]),
- as soon as a device is planned to receive an acknowledgment, no overhead is added neither in terms of protocol nor extra bits to be put into the LoRaWAN frames in uplink or downlink.

## II. REINFORCEMENT LEARNING

We model the IoT wireless spectrum issue as a Multi-Armed Bandit (MAB) problem and we propose to use bandit algorithms at the IoT device side to solve this issue.

*A. System model*

We consider the system model presented in Fig. 1, where a set of object sends uplink packets to the network gateway.

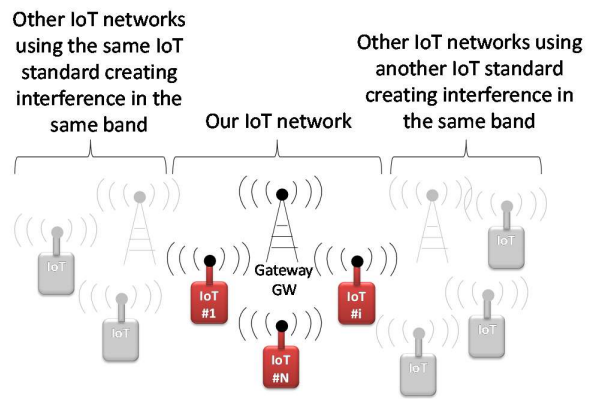

Fig. 1. System model used for IoT, with intelligent IoT devices that are able to dynamically set their transmission channel thanks to a learning algorithm, in order to minimize collisions and interference from other radio signals in the unlicensed ISM band, especially other IoT networks which will be responsible of most of future traffic.

The communication between IoT devices and this gateway is done through a simple pure ALOHA-based protocol, where devices transmit uplink packets of fixed duration whenever they want. The devices can transmit their packets in $K \geq 2$ channels. In the case where the gateway receives an uplink in one channel, it transmits an acknowledgement to the end-device in the same channel, after a fixed delay. These communications operate in unlicensed ISM bands, and consequently, as stated in previous section, suffer in particular from interferences generated by uncoordinated neighboring networks. This interfering traffic is uncontrolled, and can be unevenly distributed over the $K$ different channels.

We consider the network from the point of view of one IoT device. Every times the end-user has to communicate with the gateway (at each transmission $t \geq 1$, $t \in \mathbb{N}$), it has to choose one channel, denoted as $C(t) = k \in \{1, \ldots, K\}$. Then, the IoT device starts to wait in this channel $C(t)$ for an acknowledgement sent by the gateway. Before sending another message (i.e., at time $t + 1$), the IoT device knows if it received or not this ACK message. For this reason, selecting the channel (or *arm*) $k$ at time $t$ yields a (random) feedback, called a *reward*, $r_k(t) \in \{0, 1\}$, being 0 if no ACK was received before the next message, or 1 if ACK was successfully received. The goal of the IoT device is to minimize its packet loss ratio, or equivalently, it is to maximize its cumulative reward, as it is usually done in MAB problems [8][9][10]:

$$r_{1\ldots T} := \sum_{t=1}^{T} r_{C(t)}(t) \qquad (1)$$

This problem is a special case of the so-called "stochastic" MAB, where the sequence of rewards drawn from a given arm $k$ is assumed to be *i.i.d.*, under some distribution $v_k$, that has a mean $\mu_k$. Several types of reward distributions have been considered in the literature, for example distributions that belong to a one-dimensional exponential family (e.g., Gaussian, Exponential, Poisson or Bernoulli distributions). Rewards are binary in our model, and so we consider only Bernoulli distributions, in which $r_k(t) \sim \text{Bern}(\mu_k)$, that is, $r_k(t) \in \{0, 1\}$ and $\mathbb{P}(r_k(t) = 1) = \mu_k \in [0, 1]$. Contrary to many previous work done in the cognitive radio field (for instance in Opportunistic Spectrum Access [2]), the reward $r_k(t)$ does not come from a sensing phase before sending the $t$-th message, as it would do for any "listen-before-talk" model. Rewards come from receiving an acknowledgement from the gateway, between the $t$-th and $t+1$-th messages. The problem parameters $\mu_1, \ldots, \mu_K$ are of course unknown to the IoT device, so to maximize its cumulated reward, it must learn the distributions of the channels, in order to be able to progressively focus on the best arm (i.e., the arm with largest mean). This requires to tackle the so-called exploration-exploitation dilemma: a player (here, an IoT device) has to try all arms a sufficient number of times to get a robust estimate of their qualities, while not selecting the worst arms too many times.

Before discussing the relevance of a MAB model for our IoT application, we present two bandit algorithms, UCB$_1$ and Thompson Sampling, which are both known to be efficient for stationary i.i.d. rewards and are shown below to be useful in our IoT application.

### B. The UCB$_1$ algorithm

A first naive approach could be to use an empirical mean estimator of the rewards for each channel, and select the channel with the highest estimated mean at each time; but this "greedy" approach is known to fail dramatically [10]. Indeed, with this policy, the selection of arms is highly dependent on the first draws: if the first transmission in one channel fails and the first one on other channels succeed, the end-user will never use the first channel again, even it is the best one (i.e., the most available, in average).

Upper Confidence Bounds (UCB) algorithms instead use a confidence interval on the unknown mean $\mu_k$ of each arm, which can be viewed as adding a "bonus" exploration to the empirical mean. They follow the "optimism-in-face-of-uncertainty" principle: at each step, they play according to the best model, as the statistically best possible arm (i.e., the highest upper confidence bound) is selected. More formally, for one IoT device, we denote by

$$N_k(t) = \sum_{\tau=1}^{t} \mathbb{1}(C(\tau) = k) \qquad (2)$$

the number of times channel $k$ was selected up-to time $t \geq 1$. The empirical mean estimator of channel $k$ is defined as the mean reward obtained by selecting it up to time $t$,

$$\widehat{\mu_k}(t) = (1/N_k(t)) \sum_{\tau=1}^{t} r_k(\tau) \mathbb{1}(C(\tau) = k) \qquad (3)$$

For UCB$_1$, the confidence term is

$$A_k(t) = \sqrt{\alpha \log(t)/N_k(t)} \qquad (4)$$

And the upper confidence bound is the sum of the confidence term and the empirical mean,

$$B_k(t) = \widehat{\mu_k}(t) + A_k(t) \qquad (5)$$

which is used by the end-user to decide the channel for communicating at time step $t + 1$:

$$C(t + 1) = \arg\max_{1 \leq k \leq K} B_k(t) \qquad (6)$$

UCB$_1$ is called an index policy. The UCB$_1$ algorithm uses a parameter $\alpha > 0$, originally $\alpha$ was set to 2 [11], but empirically $\alpha = 1/2$ is known to work better (uniformly across problems), and $\alpha \geq 1/2$ is advised by the theory [11]. This algorithm is simple to implement and to use in practice, even on embedded micro-processors with limited computation and memory capabilities. In our model, every IoT device implements its own UCB$_1$ algorithm, independently. For one IoT device, the time $t$ is the total number of sent messages from the beginning, as rewards are only obtained after a transmission.

### C. The Thompson sampling algorithm

Thompson Sampling (TS) [8] was introduced early on, in 1933 as the very first bandit algorithm, in the context of clinical trials (in which each arm models the efficacy of one treatment across patients). Given a prior distribution on the mean of each arm, the algorithm selects the next arm to draw based on samples from the conjugated posterior distribution, which for Bernoulli rewards is a Beta distribution.

A Beta prior Beta($a_k(0) = 1$, $b_k(0) = 1$) (initially uniform) is assumed on $\mu_k \in [0, 1]$, and at time $t$ the posterior is Beta($a_k(t)$, $b_k(t)$). After every channel selection, the posterior is updated to have $a_k(t)$ and $b_k(t)$ counting the number of successful and failed transmissions made on channel $k$. More precisely, if the ACK message is received, the update is $a_k(t + 1) = a_k(t) + 1$, and $b_k(t + 1) = b_k(t)$, otherwise the update is $a_k(t + 1) = a_k(t)$, and $b_k(t + 1) = b_k(t) + 1$. Then, the decision is done by sampling an index for each arm, at each time step $t$, from the arm posteriors: $X_k(t) \sim \text{Beta}(a_k(t), b_k(t))$, and the chosen channel is simply the channel $C(t + 1)$ with highest index $X_k(t)$. For this reason, Thompson Sampling is called a randomized index policy.

The TS algorithm, although being simple and easy to implement, is known to perform well for stochastic problems, for which it was proven to be asymptotically optimal [12][13]. It is known to be empirically efficient, and for these reasons it has been used successfully in various applications, including on problems from Cognitive Radio [14][15], and also in previous work on decentralized IoT-like networks [16].

*D. Multi-player bandit issue*

We can prove that one single intelligent IoT can improve consequently its performance in LPWAN IoT networks using unlicensed band. But we have also shown that even if there are a lot of intelligent IoT devices, and the model of other surrounding IoT devices does not stay purely stochastic, learning still brings improvement [6]. Further theoretical developments on this direction are an interesting future work.

### III. IoT Proof-of-Concept

We first developed a proof-of-concept named MALIN, demonstrating the feasibility of using learning algorithms on the IoT device side, on real radio signals in lab conditions [4].

*A. PoC setup*

This PoC is based on 4 USRP platforms from Ettus Research and NI[1]. The development has been done using the GNU Radio[2] software, and the source code of the PoC can be found on-line[3] in order to reproduce our results. We have not implemented a real IoT standard in this PoC, in order to show that it can be applicable for any IoT standard. However, we took some characteristics rather corresponding to the LoRa context (not ultra-narrow band, reduced number of channels, frame duration around a few hundreds of milliseconds, etc.).

One USRP platform is a traffic generator which emulates as much (random) IoT traffic as we want, to be able to tune each channel's load independently, on demand. We typically choose channel loads from 0% to 20%, which is the scale supported in theory by a pure ALOHA channel access scheme.

One or two USRP platforms are playing the role of IoT devices that can run (or not) the proposed learning algorithms. They transmit at their own initiative some very light modulated information (using QPSK) so as to be identified by the gateway and then wait during one second for the gateway ACK. Both uplink transmissions and downlink receptions are done on the same frequency channel. Whether the ACK is received or not, the learning algorithm updates its knowledge about the channel used during this iteration.

A fourth USRP platform is a gateway (GW) that is continuously scanning the IoT traffic composed of the artificial signals produced by the traffic generator and the IoT platforms signals. The gateway has the ability to answer to the IoT devices, while sending back to them an ACK message containing their identifier, which is the symbols corresponding to the QPSK complex conjugate of their identifier indeed.

In order to simplify the radio signal reception we use an artificial carrier synchronization between all USRP platforms, using an Ettus Octoclock[1]. However, a simple carrier recovery method could be used. Consequently, we just have a phase correction to implement at both gateway and IoT receiver sides, from the radio point of view.

*B. PoC results*

The number of IoT channels is a parameter, and we have set it to 4, 8 and 16 channels in our experimients, but there is no limitation. For the of clarity in the figures, we give examples below with 4 channels that are separated by empty channels, but they could be contiguous with no change neither in the implementation nor in the results.

We can see on Fig. 2 a time-frequency waterfall view captured by the gateway, where we can observe the RF traffic in 4 channels. Time is vertical and going down and frequency is on the x axis. The difference of colors is a difference of received power, due to the distance of transmitters to the gateway receiver antenna. The gateway transmitter antenna is very close so signals transmitted by the gateway are red. The traffic generator and IoT devices are a little bit further away, so the gateway received weaker signals from them, one is blue and the other green, which reveals a low difference.

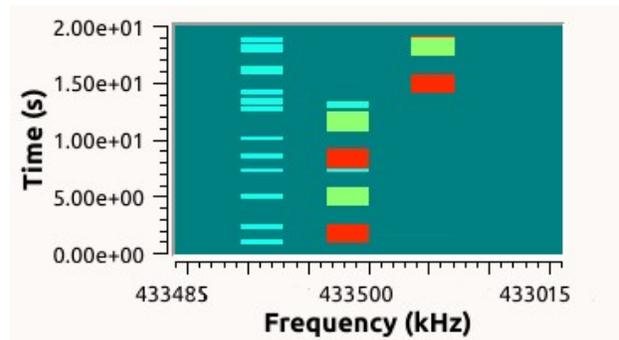

Fig. 2. Spectrum waterfall on GRC received at gateway side in a 4 channels example, during exeriments. Time is in y axis (going down) and frequency in x axis. Blue short transmissions are those produced by the traffic generator, green blocks are our IoT transmissions and red blocks are the gateway transmissions itself.

In this experiment, we can see en Fig. 2 that channel 1 has been configured to have a dense IoT traffic, which appears as blue short transmissions (produced by the traffic generator). Some others appear on channel 2, but we do not see any blue short messages on channel 3 and 4. However, we see on these channels longer messages of two kinds: green messages which correspond to IoT devices transmissions. In order to rapidly have results on the demo, we make them transmit every 5 seconds, for a message of duration of one second. Then when an IoT device transmits a message, the gateway should answer and sends an ACK to the IoT device within 1 second if the gateway was able to demodulate the signal, if there is no collision in the radio channel: these correspond to the red blocks in Fig. 2. For instance, we can see in this screenshot that the IoT device moved from channel 3 to channel 2, and at each transmission of the IoT device, the gateway was able to answer, successfully sending an ACK response.

Fig. 3 gives the perspective of the IoT device, at a different moment for the same scenario. Then we observe that colors have changed, as the received power is now device-centric. The IoT device transmitter antenna is now very close, so signals transmitted by the IoT device are red. The traffic generator, the other IoT devices, and then the gateway all are a little bit further away, so the IoT device received weaker signals from all of them, one is blue and the other green but inversed. However, it is not so obvious, so it is better to consider the message duration in the y axis indeed.

---

[1] https://www.ettus.com/
[2] https://www.gnuradio.org/
[3] https://bitbucket.org/scee_ietr/malin-multi-armed-bandit-learning-for-iot-networks-with-grc is released publicly under the open-source GPLv3 license.

We can see on Fig. 3 that if we use the same scenario of traffic as in Fig. 2, with a very dense traffic on channel 1, less dense on channel 2, even less dense on channel 3, then transmission appears on channel 4 but it is indeed just even less dense. At that time of the experiment, our IoT device is moving from channel 3, where maybe it faced some collisions in the dowlink transmission of ACK, to channel 4, where several successive transmissions and receptions seem to occur.

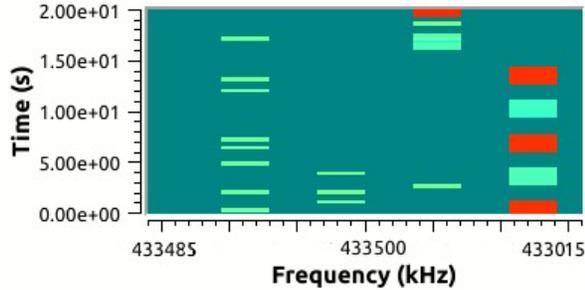

Fig. 3. Spectrum waterfall on GRC received at IoT device side in a 4 channels example, during exeriments. Time is in y axis (going down) and frequency in x axis. Green short transmissions are those produced by the traffic generator, red blocks are the IoT device transmissions, and green blocks are gateway transmissions.

Fig. 4 is a screenshot taken at some moment during an experiment, that gives the details of the learning algorithm operation. We can see in top-left red data the number of times each channel has been used. There is a clear desequilibrium with channel 4 that has been much more (17 times) used than channel 3 (8 times), itself more used than channel 2 (6 times) and channel 1 (only once). This reveals the effect of the learning algorithm. It has analyzed which channels are more occupied and disturbed by other users of the band (emulated here by the traffic generator). The top-right green and as a consequence the bottom-right blue data explain such a choice. Channel 4 has known 16 successes (over 17), so a rate of 94%. Successes mean that the IoT device received on that channel 16 ACK from the GW after transmitting 17 times in this channel. So just one "exchange" was lost, either in UL, or in DL, due to a collision with some interferring signal in the channel. We can see on the opposite that no success has been obtained for channel 1, so it has a 0% rate. UCB data, in bottom-left green part, are harder to follow, as UCB indexes rapidly converge to very close values, but at each transmission, the IoT device chooses channel with highest UCB index, as in (6).

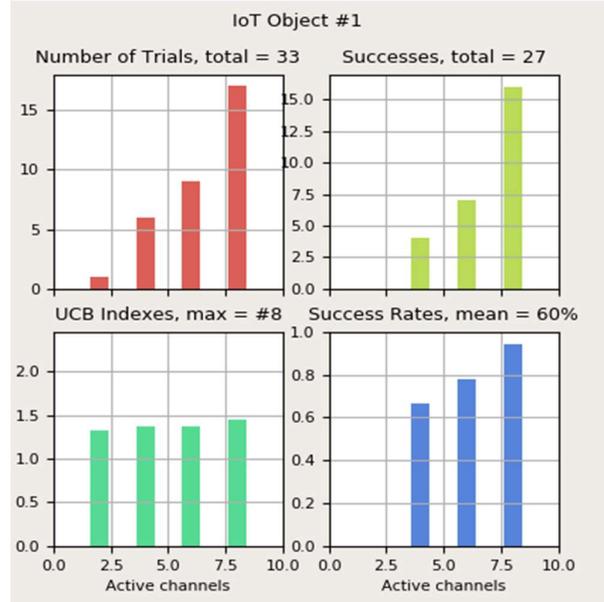

Fig. 4. Live results enabling to monitor the learning algorithm evolution at the IoT device side in a 4 channels example. Top-left red: number of trials on each channel, top-right green: number of successes on each channel (ACK received by IoT device), bottom-left green: UCB index B for each channel, bottom-rignt blue: success rate on each channel.

## IV. IoTligent: learning for non coordinated and decentralized IoT dynamic spectrum access

The next and last step after modeling and exposing our proof-of-concept consists in implementing our approach in real conditions of operation, that is, in a real IoT network. As far as the authors know, this is the first implementation of decentralized artificial intelligence algorithms in IoT devices to tackle the IoT spectrum contention mitigation problem and we named it IoTligent [20]. It is necessary, first, to remind quickly how a LoRaWAN network is constituted. We are using here a real LoRa network in the European ISM band, at 868 MHz.

### A. LoRaWAN architecture

The implementation of the learning algorithm we propose is decentralized, it takes place only on the LoRa device side. As stated earlier, no aspect of the LoRaWAN network is impacted. We explain below a little bit more the LoRaWAN network side configuration [1][19].

A LoRaWAN network, as any other IoT network, can be summarized by four main elements, as shown in Fig. 5:

- LoRa devices (our devices run the UCB algorithm here),
- LoRa gateway(s) receiving all LoRa radio signals in their radio range,
- A Lora Network Server (LNS) that discriminates devices subscribing to its network from others,
- An Application Server (AS) that receives the data sent by devices and sends back ACK to them (mandatory here).

The IoT devices are associated to a given LoRaWAN network with a "join phase", at their very first communication through a gateway of this network. The appaiment is done at the LoRa Network Server (LNS) side as explained below. Finally, data extracted from radio signals, sent by the IoT devices, are sent to the Application Server (AS) that manages data (i.e. processes them, sends them to a storing place in the cloud and/or an application). Then the role of the AS is to initiate a sending of an ACK to the IoT device, through LNS and a gateway, down to the IoT device.

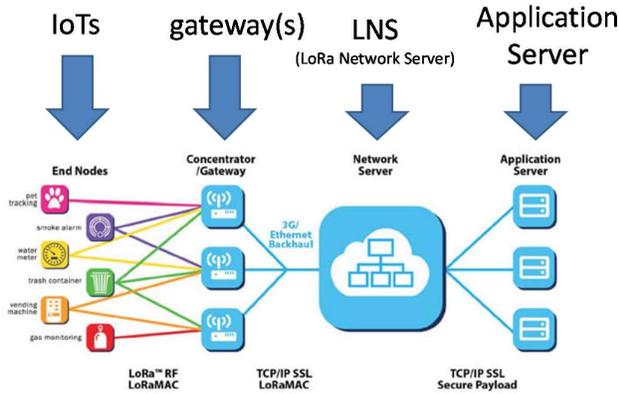

Fig. 5. LoRaWAN network parts: IoT devices, gateways, LNS and AS [19].

*B. Device side*

As an IoT device, we use, as shown on Fig. 6, a Pycom card [17] composed of an Expansion Board and a LoPy module which can support LoRa wireless connectivity. The Pycom card is programmed in the Python language. The frequency channels used in the experiments are those authorized in the country of experimentation, i.e., France. Three channels are usually used in Europe for uplink (UL) with a duty cycle of 1%: 868.1 MHz, 868.3 MHz, 868.5 MHz. Our proposal IoTligent is completely agnostic to the number of channels in the standard, so it can be used in any country.

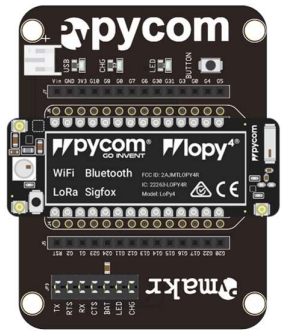

Fig. 6. Pycom module composed of a LoPy4 and an Expansion Board [17].

Note that some amendment had to be done to the Pycom firmware in order to enable programming channel assignement in the Python program, in *LORAWAN* mode.

*C. Network side*

We have access to the LNS provided by the Acklio Company. Acklio has several gateways in the city of Rennes, where the experiments were made. LNS sends the received messages to an AS which is a Linux server, running in the cloud. AS is running a Python program that enables to display data and metadata (i.e., frequency, time of reception, etc). This programs also contains instructions to send an acknowledgment to the device, using in DL the same frequency used by the IoT device at UL.

## V. IMPLEMENTATION OF IOTLIGENT IN A REAL LORA NETWORK

*A. Device side*

Based on on-line examples found in [17] we use *LORAWAN* mode with an Over-The-Air-Activation (OTAA) using *app_EUI* and *app_key* keys:

```
lora = LoRa(mode=LoRa.LORAWAN,region=LoRa.EU868)
lora.join(activation=LoRa.OTAA,   auth=(app_EUI, app_key), timeout=0)
```

The transmit channel frequency is then chosen in a set of N channels which is set here at N = 3. We use standard Europe UL channels with the following frequency table (in MHz):

```
tabFreq =[868100000, 868300000, 868500000]
```

IoTligent device infinite *while loop* is started, running the algorithm presented in the next section and [2], in order to choose which frequency to be selected at each iteration before executing a send operation. An ACK is then expected from the network side in *non blocking* mode so that when ACK is not received, device just updates its learning data and still goes on.

*B. Network side – Lora Network Server*

Devices should be declared to LNS with at least the following information:

- *devEUI* : ID of the device obtained by executing a « *get_id.py* » program from [17] on the Pycom device itself,
- *appEUI* : which should correspond to *app_eui* chosen in the pycom device,
- *appKey*: which should correspond to *app_key* chosen in the pycom device,
- other parameters are let by default at SF=12 (spreading factor), and bandwidth BW=125kHz.

The address of the AS is also specified in *Connectors*, as well as the mode used to send data between LNS and AS (*http callback* chosen here).

*C. Network side – Application Server*

The AS runs a Python program that receives data from the LNS, as well as LoRa metadata with all parameters of LoRaWAN transmission (frequency, SF, BW, time of arrival, etc). This program also sends an acknowledgment message to the device in DL. First, an acknowledgment attempt is sent by default at the same frequency than the message transmitted by device it answers to. Then we block any other retransmission. This is exactly what is necessary for the learning of IoTligent:

- to use the same channel in both UL and DL,
- to avoid retransmission in order to save batteries of devices on the one hand, and radio frequency overload on the other hand.

## VI. LEARNING ALGORITHM IN PYCOM DEVICE

The learning algorithms used in IoTligent are (any) bandit algorithms, such as those first used for Cognitive Radio dynamic spectrum access in [2], and implemented in the SMPyBandits Python library [18]. We take here the example of $UCB_1$ algorithm, as presented above [5]. We have chosen these algorithms for their ease of implementation. The only data necessary to be stored for $UCB_1$ algorithm are:

- an iteration index initialized at 0: `itindex`,
- a table of size N (the number of channels, 3 in this implementation example, but it could be arbitrarily high) for the number of times each channel has been chosen, representing $N_k$ of (2): `Tk[]`.
- another table of size N for the empirical mean of success of each channel, i.e., $\widehat{\mu_k}(t)$ of (3): `Xk[]`.

From the learning algorithm point of view, a success occurs when an IoT device receives an ACK from the IoT network (as explained above), which means that the currently used frequency channel suffered no collision both in UL and DL. Otherwise, a failure occurred. The update of the selected channel empirical mean $X_k$ is reconstructed easily from the number of activations and previous $X_k$ stored value. So it is not necessary to store in memory the results of all past iterations, but just only a summary of it (its mean).

Then after an initialization phase where each channel is selected alternatively once, UCB$_1$ algorithm really starts [2] . It consists for each iteration in choosing the frequency channel with greatest index $B_k$ as defined in (5), with bias $A_k$ of (4), that is computed for each channel like this in a *for loop* on i index, and with *alpha* the UCB$_1$ parameter $\alpha$ that sets the exploration vs. exploitation trade-off [2]:

`Ak[i] = math.sqrt(alpha*math.log(it)/Tk[i])`

IoTligent channel selection is then on the greatest $B_k$ [2]:

```
for i in range(N):
  Bk[i] = Xk[i] + Ak[i]
  if Bk[i] > bestChannel:
     bestChannel = Bk[i] ; freq = tabFreq[i]
```

## VII. RESULTS FOR THE IOTLIGENT DEMONSTRATION

Experiments have been done on a real LoRa network currently deployed with 3 channels. More channels are expected to be used in the future, but it will not induce any implementation difference (only 2 extra numbers to be saved by added channel). We present results obtained on a IoTligent device, for 129 transmissions done every 2 hours, so a period of 11 days. Fig. 7 shows the evolution of the *Tk* index through time, the number of time each channel has been selected by the learning algorithm. In the figures, the black curve is for channel 0 (at 868.1 MHz), the blue curve is for channel 1 (868.3 MHz) and the red curve is for channel 2 (868.5 MHz).

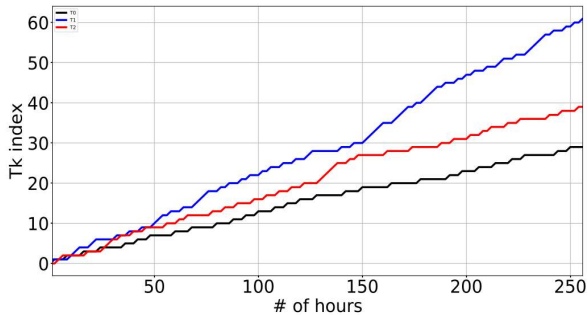

Fig. 7. Evolution of the *Tk* index through time (as learning happens).

Fig. 8 gives the empirical mean *Xk* experienced by the device on each of the 3 channels. Each peak corresponds to a successful LoRa bi-directional exchange between the device and AS: from device uplink transmission to ACK reception (downlink) by the device.

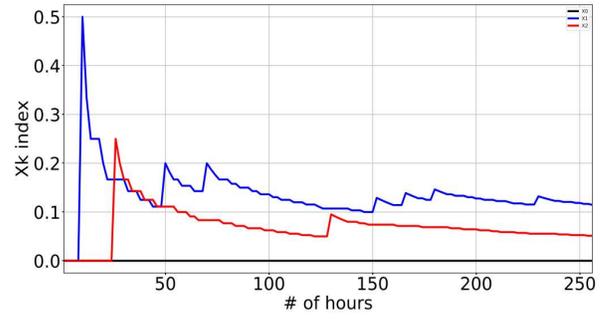

Fig. 8. Evolution of the *Xk* empirical mean through time. Black curve: channel 0 ; blue curve: channel 1 ; red curve: channel 2.

We can see that channel 1 gives the best results, before channel 2, but channel 0 always failed in sending back an ACK to the device. Each peak in Fig. 8 reveals a successful case where ACK has been received by IoTligent device. Fig. 8 gives the end results after 11 days. We can see that channel 0 has been tried 29 times with *Sk[0] = 0* success (i.e., no ACK received by the device). So the learning algorithm made the device use 61 times channel 1 with *Sk[1] = 7* successful bi-directional exchanges, and 39 times channel 2 with *Sk[2] = 2* successes. This corresponds to 7 (respectively 2) peaks of *Xk[1]* (respectively *Xk[2]*) on Fig. 2.

TABLE I. RESULTS AT THE END OF THE EXPERIMENT

| Tk[0] = 29 | Tk[1] = 61 | Tk[2] = 39 |
|---|---|---|
| Xk[0] = 0.0 | Xk[1] = 0.115 | Xk[2] = 0.051 |
| Sk[0] = 0 | Sk[1] = 7 | Sk[2] = 2 |

The empirical mean gives the vision the device obtained from the channels, i.e., a mean probability of 11.5% of successful bi-directional connection for channel 1 and 5% for channel 2, whereas channel 0 never worked from the device point of view. With a normal device, i.e., a non IoTligent device, a random access is done, trying once over 3 times on each channel, for a global average successful rate of 5.5%.

It is important to note that here the learning algorithm is mostly in its exploration phase, but is learning very fast. Only during the last 2 days of the experimen, channel 1 has already been used 4 times more than channel 0 and 2.5 times more than channel 2, which means that learning is already effective. As proven for UCB algorithms [2][3], channel 1 will be more and more selected so that the global success rate will converge to the percentage of success of the best channel, which is 11.5% in this experiment (this estimate can be considered as a good evaluation as it is based on 61 trials). In other words, this means that a mean of 15 successes can be expected in the long term over the same period of 11 days with IoTligent. On the contrary, normal devices will never improve and stay in the current average, i.e. in average 7 successful transmission on the same period duration.

In order to have the same rate of successful transmissions, normal IoT devices should consequently transmit twice more often, which has two negatives impacts. The first impact is that normal IoT devices autonomy will be twice less than IoTligent devices. The second but not the least impact is that devices will occupy twice more times radio channels, hence contributing to increase even more the risks of radio collisions and thus the IoT bands congestion.

## VIII. Conclusion and Perspectives

We describe in this paper the solution we propose to mitigate radio collisions in IoT unlicensed bands. Our solution is based on learning algorithms to be implemented on the IoT device side, at a very low cost of implementation and no protocol overhead. We prove the efficiency of the method on a proof-of-concept demonstration based on USRP platforms in laboratory conditions (named MALIN), then we present the implementation of learning algorithms on devices deployed in a real IoT network. Implementation on LoRa devices in a real LoRaWAN network is demonstrated and is named IoTligent. As far as we know, we propose the first implementation of a decentralized spectrum learning scheme for IoT wireless networks. Even if the current IoT networks are (yet) not densely populated by devices, medium and even short term forecasts predict a high number of devices to overcrowd ISM unlicensed bands. The IoTligent approach is then a solution to extend IoT devices battery life, which is a key performance indicator in any IoT eco-system.

## Acknowledgment

The authors would like to thank Rémi Bonnefoi for the MALIN implementation [4], as well as Laurent Toutain, from IMT Atlantique and the Acklio Company, and Yalla Diop for their technical support on LoRa network and Pycom programming.

## References

[1] N. Sornin, M. Luis, T. Eirich and A. L. Beylot "LoRaWAN specification", technical report, LoRa Alliance, Inc., January 2015.

[2] W. Jouini, D. Ernst, C. Moy and J. Palicot, "Upper Confidence Bound Based Decision Making Strategies and Dynamic Spectrum Access," *IEEE ICC, International Conference on Communications,* Cape Town, South Africa, May, 2010.

[3] C. Moy, "Reinforcement Learning Real Experiments for Opportunistic Spectrum Access", *Karlsruhe Workshop on Software Radio*, Karlsruhe, Germany, March 2014.

[4] L. Besson, R. Bonnefoi, C. Moy, *"MALIN: Multi-Armed bandit Learning for Iot Networks with GRC*: A TestBed Implementation and Demonstration that Learning Helps", ICT 2018, France, June 2018.

[5] P. Auer, N. Cesa-Bianchi, and P. Fischer, "Finite-time analysis of the multiarmed bandit problem", Machine Learning, volume 47, number 2-3, May 2002.

[6] R. Bonnefoi, L. Besson, C. Moy, E. Kaufman and J. Palicot, "Multi-Armed Bandit Learning in IoT Networks: Learning helps even in non-stationary settings", CROWNCOM 2017, Lisbon, September 2017.

[7] A. Anandkumar, N. Michael, A. K. Tang, and A. Swami, "Distributed algorithms for learning and cognitive medium access with logarithmic regret", IEEE J. Sel. Areas Commun., v. 29, no. 4, Apr. 2011.

[8] W. R. Thompson, "On the likelihood that one unknown probability exceeds another in view of the evidence of two samples," Biometrika, vol. 25, 1933.

[9] H. Robbins, "Some aspects of the sequential design of experiments," Bulletin of the American Mathematical Society, vol. 58, no. 5, pp. 527–535, 1952.

[10] T. L. Lai and H. Robbins, "Asymptotically efficient adaptive allocation rules," Advances in Applied Mathematics, vol. 6, no. 1, pp. 4–22, 1985.

[11] S. Bubeck and N. Cesa-Bianchi, "Regret analysis of Stochastic and Non-Stochastic Multi-Armed Bandit Problems," Foundations and Trends® in Machine Learning, vol. 5, no. 1, pp. 1–122, 2012.

[12] S. Agrawal and N. Goyal, "Analysis of Thompson sampling for the Multi-Armed Bandit problem", in JMLR, Conference On Learning Theory, 2012.

[13] E. Kaufmann, N. Korda, and R. Munos, "Thompson Sampling: an Asymptotically Optimal Finite-Time Analysis", pp. 199–213. Springer, Berlin Heidelberg, 2012.

[14] V. Toldov, L. Clavier, V. Loscrí and N. Mitton, "A Thompson Sampling approach to channel exploration-exploitation problem in multihop cognitive radio networks", in PIMRC, 2016.

[15] A. Maskooki, V. Toldov, L. Clavier, V. Loscrí, and N. Mitton, "Competition: Channel Exploration/Exploitation Based on a Thompson Sampling Approach in a Radio Cognitive Environment", EWSN, 2016.

[16] C. Moy, J. Palicot, and S. J. Darak, "Proof-of-Concept System for Opportunistic Spectrum Access in Multi-user Decentralized Networks", EAI Endorsed Transactions on Cognitive Communications, volume 2, 2016.

[17] Pycom documentation: https://GitHub.com/PyCom/PyCom-libraries

[18] L. Besson, "SMPyBandits: an Open-Source Research Framework for Single and Multi-Players Multi-Arms Bandits (MAB) Algorithms in Python": https://GitHub.com/SMPyBandits/SMPyBandits https://SMPyBandits.GitHub.io/

[19] LoRaWAN™ v1.1 Specification, 2017, LoRa Alliance Inc, https://LoRa-alliance.org/sites/default/files/2018-04/lorawantm_specification_-v1.1.pdf

[20] C. Moy, "IoTligent: First World-Wide Implementation of Decentralized Spectrum Learning for IoT Wireless Networks", URSI AP-RASC, New Delhi, India, 9-14 March 2019.